\documentclass[runningheads]{llncs}
\usepackage[T1]{fontenc}
\usepackage{graphicx}
\begin{document}
\title{Modeling and Analysis of the Lead-Lag Network of Economic Indicators} 
%
%
\author{Amanda Goodrick\inst{1,2}\orcidID{0000-0001-8555-590X} \\
\and Hiroki Sayama\inst{1,3}\orcidID{0000-0002-2670-5864}}
\authorrunning{A. Goodrick et al.}
%
\institute{State University of New York at Binghamton, Binghamton NY 13902, USA \and
Slippery Rock University, Slippery Rock PA 16057, USA 
\email{agoodri2@binghamton.edu}\\
\and
Waseda Innovation Lab, Waseda University, Tokyo, Japan\\
\email{sayama@binghamton.edu}}
\maketitle              
\begin{abstract}
We propose a method of analyzing multivariate time series data that investigates lead-lag relationships among economic indicators during the COVID-19 era with a weighted directed network of lagged variables. The analysis includes a stock index, average unemployment, and several variables that are used to calculate inflation. Three complex networks are created with these variables and several lags of each as the network nodes. Network edges are weighted based on three relationship metrics: correlation, mutual information, and transfer entropy. In each network, nodes are merged, and edges are aggregated to simplify the weighted directed graph. Pagerank is used to determine the most influential and the most influenced node over the time period. Results were reasonably robust within each network, but they were heavily dependent on the choice of metric.

\end{abstract}
\section{Introduction}
Economic activity is relevant to everyone; it impacts daily choices and determines lifestyles. The economy has been quite dynamic in recent years, particularly throughout the COVID-19 era. Understanding the dynamics of some of the underlying economic factors can help us to make choices that will improve our quality of life. Fluctuations in the stock market have historically been thought to indicate a future change in the economy, but those fluctuations are frequent and unpredictable. The unemployment rate tends to vary within a reasonable range, and it has been considered to indicate past changes in the economy, as a poor economy leads to fewer jobs. With fewer jobs, less money is circulating in the system, and the economy is further impacted by the change in the value of a basket of goods, impacting inflation. Relationships among these indicators can be investigated by considering the lead-lag effect in this system. A lead-lag effect in time series is a relationship between lead variables and lagged variables, where lagged variables are the variables shifted backward in time. A lead variable is a variable that happens first and has a relationship with other variables at some future time, and it provides information about future system behavior. A lagged variable is a variable that has a relationship with other variables at some past time, and it provides information about past system behavior. The stock market has historically been considered as a lead indicator that will predict future activity \cite{stock lead,stock lead2,GC}. Unemployment rate and inflation have both been considered as lagged indicators that are the result of the system state \cite{stock lead,stock lead2,URStockInfl}. We typically consider each of these indicators to determine overall economic health, but how do they each interact? To be better active consumers, we must understand the lead-lag relationships and the most relevant variables in this system, particularly as those factors pertain to our everyday basic needs.

Historical approaches to infer influential relations among time series indicators involve using Granger causality, which infers an influential relationship over time using lagged variables. Previous relevant research indicates that there is a Granger causality from the stock market to the economy, though the relationship is not reciprocated \cite{GC}. That bivariate study used GDP (Gross Domestic Product) to measure the economy as it relates to the S\&P 500. Another study focused on multivariate Granger causality between financial development, investment, and economic growth and found a difference in results in the short-term vs. the long term \cite{GCM}. Neither of these studies used a network model. Multivariate Granger causality results are similar to what we want to investigate, and literature is rich in this topic, but most studies do not incorporate a network analysis, and Granger causality requires stationarity to fit a VAR model to linear time series. Granger causality has been extended to a method to detect nonlinear causality, and it has been found that transfer entropy can be used in place of Granger causality for Gaussian processes \cite{GCTE,NGC}. Other studies use Granger with sliding windows over time to detect nonlinear relationships in each window of time \cite{Jiang,Sandberg}. However, Granger causality indicates the existence of a pairwise relationship between original variables, but it does not provide an estimation of the strength of the relationship. As network analysis has become more prevalent, networks with edge weights based on correlation have commonly been used to represent the relationship strength between two variables. In much of the literature, when time series are represented by correlation networks, the goal is to detect community structure over the entire time period \cite{cor}. Recently, network clustering methods were used to detect lead-lag relationships in stock indices \cite{journal3}. Other current methods are based on constructing multilayer temporal networks with instantaneous correlations as intralayer edges \cite{multilayer}. Almost all network studies that use correlation as an edge weight are based on only positive correlations above a certain threshold in order to use only statistically significant correlations, including one recent study that investigates lead-lag correlations \cite{emergence}. However, even low or negative correlations are a measure of a pairwise relationship. As Information Theory has developed, mutual information has been used instead of correlation to quantify nonlinear relations, since correlation is intended for linear relations. Transfer entropy has recently become more popular as a connection method, and a multivariate transfer entropy has been proposed with clustering methods to infer network properties \cite{MultiTE}. Transfer entropy has been used for network connections to rank financial nodes based on Random Forest and Gradient Boosting \cite{TErandom,TErandom2}. Pagerank (PR) can be used for ranking, as it is an iterative method intended to detect the most important nodes in a network. It is based on incoming edges, so it is most appropriate for directed networks \cite{PR}. PR has previously been used to extract features in climate data using lagged positive correlations with a threshold, and it has been used with statistically significant correlations of one-minute stock log returns to rank foreign exchange markets \cite{journal1,journal2}. 

This model analysis investigates the relationship between economic indicators during the COVID-19 era using a weighted directed network analysis approach to investigate the lead-lag effect in this system. PR is used to determine the most influential and the most influenced node in a directed correlation network of lagged variables using three different relationship metrics: linear correlation, mutual information, and transfer entropy. Variables used to represent the system were several variables that are used in the calculation of Consumer Price Index (CPI), which is a measure of inflation, average unemployment rate, and the S\&P 500 stock index. We investigate the best relationship metric choice, the most influential variables, and the most influenced variables in this system. To our knowledge, there is no study that investigates this phenomenon with a complex network that retains and incorporates relationship strength with time difference and a damping parameter. For each network, the graph is simplified, while retaining all relational information, and PR is used. This novel method can be applied to multivariate time series in any field to determine lead-lag relationships, better understand time series systems, and make more informed decisions.   

\section{Methodology}

We used historical monthly data of $13$ economic variables from July 2019 to December 2022 to investigate the economic system during the COVID-19 era. The data was from the Northeast Region of USA, as classified by the Bureau of Labor Statistics \cite{data1}. This region includes 9 states and Puerto Rico. The first $11$ variables were amount spent on: apparel, food at home, food away from home, alcohol, housing, medical, recreation, new cars, used cars, motor fuel, and fuels \& utilities \cite{data1}. Average unemployment rate for the region was calculated from monthly unemployment rate per state, and monthly closing prices for the S\&P 500 stock index were obtained from Yahoo! Finance \cite{data1,data2}. This stock index was chosen because it is often considered to be an accurate overall market gauge since it is value-related as opposed to price-related \cite{stock lead2}.  

The original data had $42$ monthly values per variable. These variables were transformed to monthly rate of change to create stationarity and remove autocorrelation, resulting in $41$ monthly rates of change. The first $12$ lags for each variable were created, which increased the number of variables to $169$, since there were $12$ new variables for each of the existing $13$ variables. Each lagged variable had a sample size one less than the previous lag, so the variable of the greatest lag had 29 monthly rate of change values.  
We let $X_{ik}$ denote the lag $k$ of variable $X_i$ for $i=1,2,...,13$ and $k=0,1,...,12$, where $k=0$ indicates the original variables. A directed graph was created with the $169$ variables as nodes with a directed edge from node $X_{ik}$ to node $X_{jm}$ if $k>m$, for $i \neq j$ and $k \neq m$. Then each existing edge was directed toward the pairwise lead variable, and instantaneous correlations and any lingering autocorrelations were neglected. Edge weights were based on: pairwise absolute simple linear correlations, 

\begin{equation}
\label{eq: corr}
|r_{ik,jm}| = \left|\frac{COV(X_{ik},X_{jm})}{\sqrt{VAR(X_{ik})VAR(X_{jm})}}\right|, 
\hspace{0.05in}\mathrm{for} \hspace{0.05in} i \neq j \hspace{0.05in}\mathrm{and } \hspace{0.05in} k \neq m,
\end{equation}
pairwise lag difference, 
\begin{equation}
\label{eq: LD}
 D_{km}=|k-m|, 
\hspace{0.05in}\mathrm{for} \hspace{0.05in} k \neq m, 
\end{equation}
and a parameter $a$ where $0 \leq a \leq 1$. This parameter represents the proportion of correlation remaining over time $D_{km}$. Then each edge weight is given by 
\begin{equation}
\label{eq: weight}
w_{ik,jm}=|r_{ik,jm}|a^{D_{km}}.
\end{equation}

This $169$-node network was simplified by aggregating nodes of each variable over all lag values. For each node $i$, $X_{ik}$ was merged into a single node for all $k$, resulting in a $13$ node graph. For each directed edge from node $i$ to node $j$, edge weights were summed so that each node pair had exactly two directed edges, each with a single weight,
\begin{equation}
\label{eq: eqtn5}
 w_{i,j}=\sum_{k}\sum_{m} w_{ik,jm}.
\end{equation}
This simplified graph had $13$ nodes and $156$ edges. Weighted edges were directed toward the lead variable, so the node with the greatest PR value was considered to be the most influential node over the time investigated. Edge weights remained the same, and the direction of each edge was reversed and directed toward the lagged variable of the pair. PR was used to determine the most influenced node during the COVID-19 era.

The PR algorithm is an iterative process originally used by Google to determine the most important websites. The PR of a node is based on the incoming edges of that node divided by the number of total edges in the graph. It is given by

\begin{equation}
\label{eq: PR}
PR_i = \frac{1-d}{N} + d \sum{\frac{PR_j}{c_j}}
\end{equation}

where $PR_i$ is the pagerank of node $i$, $d$ is the damping factor, $PR_j$ is the PR value for all other nodes in the network where $j \ne i$, and $c_j$ is the number of outgoing edges from node $j$ for $j \ne i$ \cite{PR}. For a weighted graph, the pagerank algorithm in the networkx package in python uses edge weights rather than counts \cite{nx}. The damping factor is based on the idea that a web surfer will eventually stop clicking links on the internet. The default value in the pagerank algorithm was $d=0.85$, which we used for this project because the overall economy dynamics do not dampen over time, and the time factor was already accounted for in the edge weights. Results were not impacted by other reasonable choices for $d$, as the variable relationships remained intact regardless of this parameter value.
 
Since Eq. \ref{eq: PR} calculates a measure of importance based on incoming links, when edges were directed backward in time toward the pairwise lead variable, the node with the greatest PR was considered to be the most influential. When edges were directed toward the pairwise lag variable, the node with the greatest PR value was considered as the most influenced variable. It is important to note that influential and influenced in no way imply causation. An influential observation in regression is one that influences the regression equation, and the regression equation will change if the influential observation is removed. This terminology is analogous to that situation.      

This process was replicated with a network that had edge weights based on mutual information, $I(X;Y)$, instead of linear correlation, where $I(X;Y)=I(Y;X)$ and 
\begin{equation}
\label{eq: MI}
I(X;Y)=H(X) + H(Y) - H(X,Y)=\int\int{f(xy)log \frac{f(xy)}{f(x)f(y)}}.
\end{equation}

In Eq. \ref{eq: MI}, $H(X)$ is defined as the entropy of a variable, $H(X,Y)$ is the joint entropy, $f(x)$ is a probability density function, and $I(X;Y)$ is the amount of information obtained about a variable $X$ by knowing about a variable $Y$ \cite{MI}. The logarithmic function is typically base $2$ and mutual information returned is in bits of information. When variables are discrete, the integral is replaced with a summation. For independent variables, $I(X;Y)=0$, and for dependent variables, $I(X;Y) > 0$, so all of the mutual information values used as edge weights were positive. 

While correlation is based on the variable outcomes, mutual information is determined by the probability that each outcome will occur, and the outcome itself is irrelevant. Correlation is based on linear relationships, but mutual information detects all relations, so it is more appropriate to use this metric when nonlinear relationships exist.   
Since the variables were continuous, to avoid variations based on binning, pairwise mutual information was calculated using the mutual\_info\_regression function in python from the sklearn.feature\_library \cite{sklearn}. This method estimates a probability density function for each continuous variable based on the distance between each outcome and it's k-nearest neighbors \cite{regress}. The algorithm calculates the distance from the observation to the farthest neighbor, and then it counts the number of unique observations within that distance to estimate a frequency for calculating the necessary probability density functions \cite{regress,mutual}. The necessary marginal probabilities hold for all values of $k$, but the joint probabilities are more sensitive to choices of $k$. In order to avoid numerical variations based on the number of nearest-neighbors between the marginal and joint probabilities, this method starts by applying the chosen $k$-value to the joint probability estimate and works backward to fit the marginal probabilities based on the nearest neighbors used for the joint probabilities \cite{TE,regress}. The default value in this function is for $k=3$ nearest neighbors, and this value was used to calculate pairwise mutual information because values of $k$ greater than $3$ may result in an estimation with less variation but perhaps more bias \cite{regress,mutual}. We tried several other values to create the network, but results were much more consistent for values of the parameter $a$ when $k=3$. Then Eq. \ref{eq: MI} was substituted into Eq. \ref{eq: weight} in place of Eq. \ref{eq: corr} to obtain edge weights, and PR was used to determine the most influential node and the most influenced node of the mutual information network.  

The method was repeated one more time using transfer entropy to quantify the pairwise relationship between variables. Transfer entropy, $T_{X->Y}$, is the amount of information gain in future values of $Y$ by knowing past values of $X$ and past values of $Y$. Transfer entropy is given by
\begin{equation}
\label{eq: TE}
T_{X \to Y}=I(Y_t^{(l)}:X_{t-1}^{(l)}|Y_{t-1}^{(k)})=H(Y_t|Y_{t-1}^{(k)})-H(Y_t|Y_{t-1}^{(k)}X_{t-1}^{(l)})
\end{equation}
where $l$ is the variable lags from $1$ through fixed maximum lag $l$, and $H(X|Y)$ and $I(X;Y)$ are conditional entropy and mutual information, respectively \cite{TE}.
Transfer entropy is similar to mutual information in that it can be used for dynamic nonlinear relations. However, it is not a symmetric relation, $T_{X \to Y} \ne T_{Y \to X}$, so the direction of the relation matters, and $T_{X \to Y} \ge 0$, so all edge weights were positive. For this network, relations are asymmetric, and edges only exist for relations directed to a future time where $k>m$, so any relations where $k<m$ were disregarded. 
Pairwise transfer entropy was calculated using the te\_compute function from the python implementation PyIF \cite{PyIF}. This algorithm calculates transfer entropy as the sum of two mutual information values, and it is based on the same density estimates that we used for calculating mutual information \cite{regress,PyIF}. Using this function requires choosing a value of $k$ for nearest neighbors in calculating the mutual information required to calculate transfer entropy, so we again used $k=3$ to more easily compare results and to minimize bias in the estimates \cite{regress,PyIF}. The fixed maximum length of $l$ was chosen as $1$ due to the dynamics of each variable and the volatility of those with much larger standard deviations relative to the others. It was also important to not overestimate the transfer entropy in the network when lagged variables were already included. 
As with Eq. \ref{eq: MI}, Eq. \ref{eq: TE} was substituted into Eq. \ref{eq: weight} in place of Eq. \ref{eq: corr}. In each case, the simplified graph had $13$ nodes and $156$ edges. 

\section{Results}
\begin{figure}\centering
\includegraphics[width=90mm]{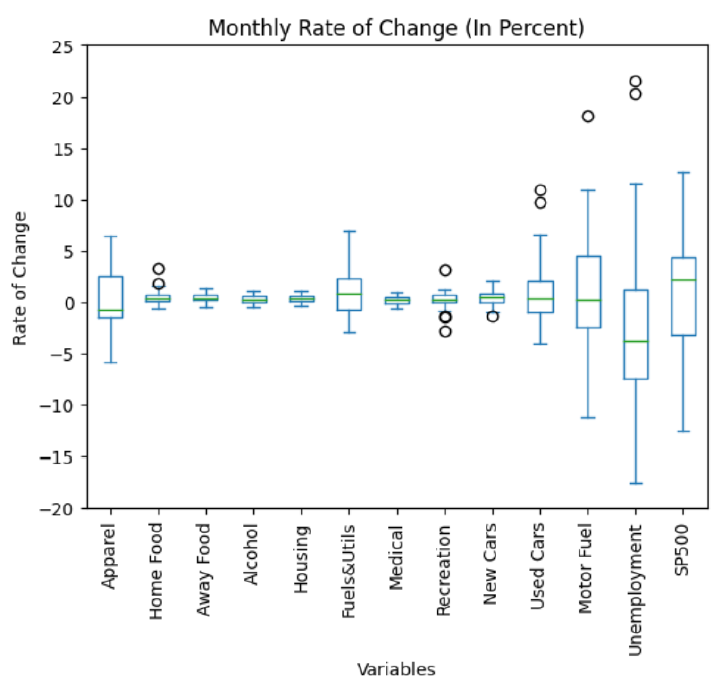} 
\caption{Rate of change (y-axis) per variable from July 2019 through December 2022. The y-axis is truncated for visibility, with an off-the-chart value of 197\% increase in unemployment from March to April 2020.} \label{fig1}
\end{figure}

The box plots in Fig. \ref{fig1} illustrate some of the basic descriptive statistics of each variable in the COVID-19 era. These descriptives are based on the monthly rate of change in percentages. Basic descriptive statistics, in percents, are provided in Table \ref{tab1}. The mean monthly rate of change is reasonably close, less than $1\%$, for each variable except apparel, whose rate of change of $0.04\%$ is relatively low, and unemployment rate, whose rate of change of $218\%$ is relatively high. There is much variation over the time period, particularly in average unemployment rate ($32.22\%$), motor fuel ($5.82\%$), S\&P 500 ($5.76\%$), and used cars ($3.18\%$). 


\begin{table}\centering
\caption{Basic Descriptives per Variable (In Percent)}\label{tab1}
\begin{tabular}{|c|c|c|c|c|}
\hline
Variable & Mean & Std. Dev. & Min & Max\\
\hline
Apparel & 0.04 & 2.95 & -5.83 & 6.38\\
Home Food & 0.47 & 0.72 & -0.62 & 3.25\\
Away Food & 0.45 & 0.41 & -0.45 & 1.35\\
Alcohol & 0.28 & 0.40 & -0.51 & 1.13\\
Housing & 0.32 & 0.36 & -0.41 & 1.03\\
Fuels \& Utils & 0.80 & 2.07 & -2.95 & 6.86\\
Medical & 0.24 & 0.42 & -0.65 & 0.94\\
Recreation & 0.20 & 0.89 & -2.77 & 3.12\\
New Cars & 0.41 & 0.81 & -1.31 & 2.09\\
Used Cars & 0.76 & 3.18 & -4.00 & 11.00\\
Motor Fuel & 0.69 & 5.82 & -11.25 & 18.10\\
Unemployment & 2.18 & 32.22 & -17.57 & 197.20\\
S\&P500 & 0.91 & 5.76 & -12.51 & 12.68\\
\hline
\end{tabular}
\end{table} 

\begin{figure}\centering
\includegraphics[width=100mm]{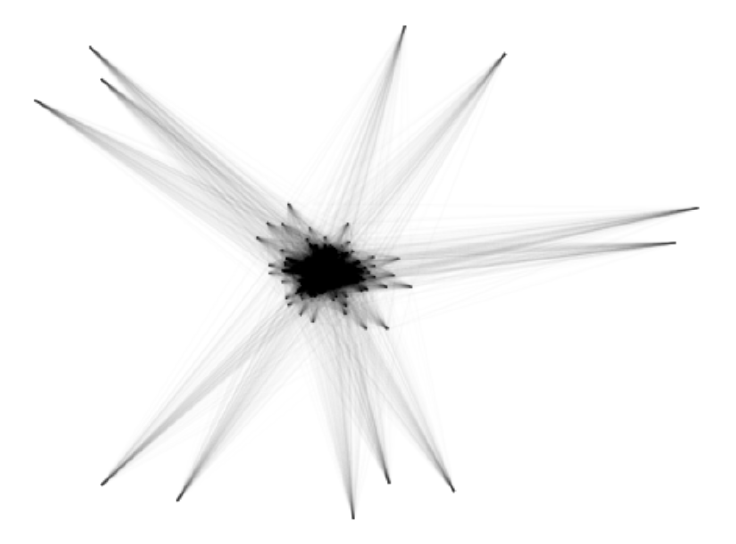} 
\caption{Correlation network. This directed network has edge weights based on Eq. \ref{eq: weight} with correlation and parameter $a=1$. This graph has $169$ nodes and $12168$ edges.} \label{fig2}
\end{figure}

\begin{figure}\centering
\includegraphics[width=100mm]{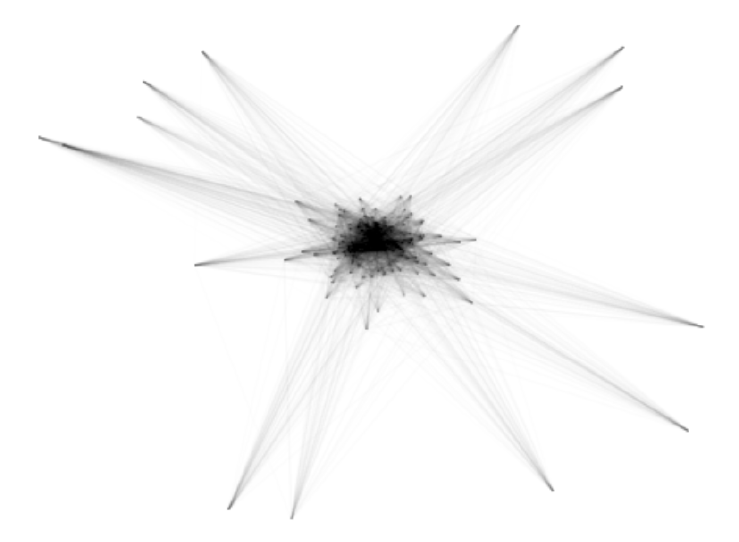} 
\caption{Mutual information network. This directed network has edge weights based on Eq. \ref{eq: weight} with mutual information Eq. \ref{eq: MI} and parameter $a=1$. This graph has $169$ nodes and $6258$ edges.} \label{fig3}
\end{figure}

\begin{figure}\centering
\includegraphics[width=100mm]{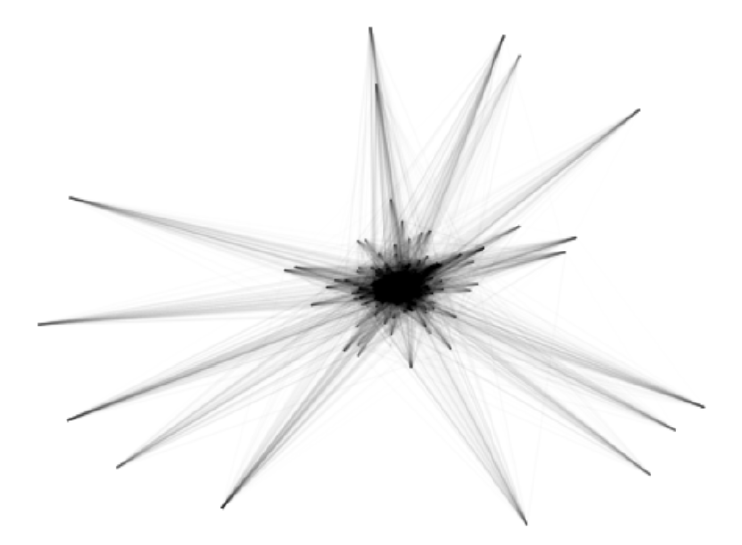} 
\caption{Transfer entropy network. This directed network has edge weights based on Eq. \ref{eq: weight} with transfer entropy Eq. \ref{eq: TE} and parameter $a=1$. This graph has $169$ nodes and $6610$ edges.} \label{fig4}
\end{figure}

The initial networks were created and are visualized in Figs. \ref{fig2}, \ref{fig3}, and \ref{fig4}, with edge weights based on correlation, mutual information, and transfer entropy, respectively. These graphs each had edges directed toward the lead variable and edge weights with parameter $a=1$. The graph with edge weights based on correlation had $12168$ edges, while edge weights based on mutual information resulted in a graph with $6258$ edges, and the graph with edge weights based on transfer entropy had $6610$ edges. The correlation and mutual information networks had edges based on all pairwise relations where Eq. \ref{eq: LD} was positive. However, while several correlation values were quite low, there were no zero correlations, so the correlation network included all possible relations. The mutual information network had fewer relations because the number of possible relations was the same as that of the correlation network, but there were pairwise relations with zero mutual information. The transfer entropy network had fewer edges than the correlation network and slightly more edges than the mutual information network. 

\begin{figure}
\includegraphics[width=\textwidth]{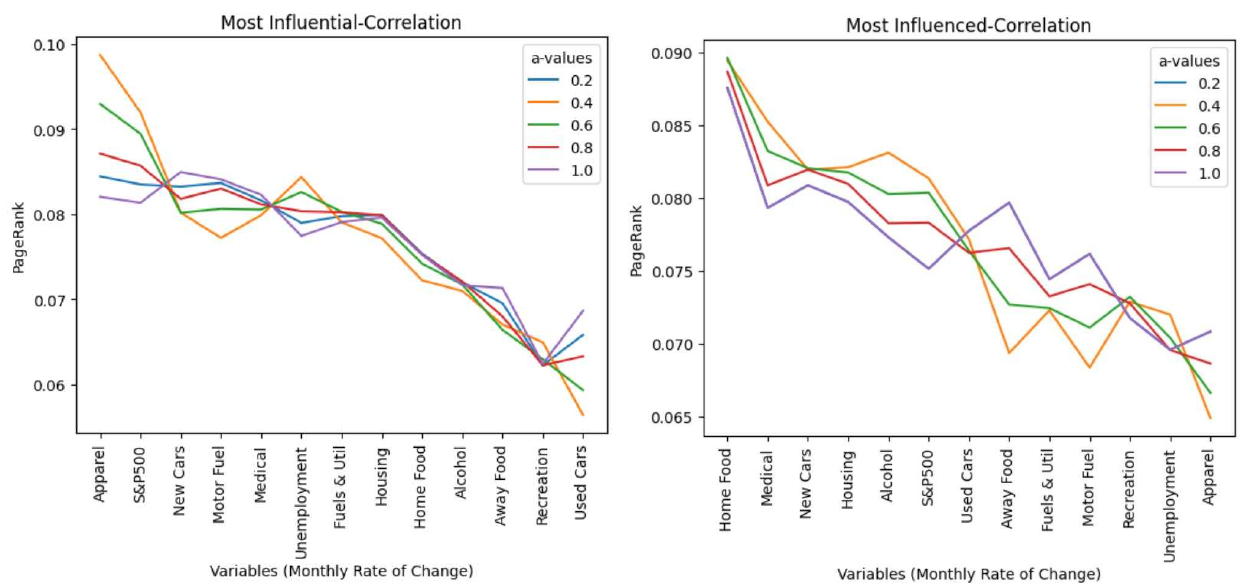} 
\caption{PR values (y-axis) per node (x-axis) for networks with edge weights based on correlation and determined by select values of $a$ in the legend. The nodes on the x-axis are sorted by average PR. The most influential node is shown on the left and has greatest PR with edges directed toward the lead variable. The most influenced node is shown on the right and has greatest PR with edges directed toward the lagged variable.} \label{fig5}
\end{figure}

Once all nodes were merged and edge weights were summed by Eq. \ref{eq: eqtn5}, PR was calculated and plotted to investigate behavior. The PR values based on the correlation network are shown in Fig. \ref{fig5}, with the most influential nodes shown on the left, and the most influenced nodes shown on the right. Variables on the x-axis are sorted by average PR, least to greatest. When edge weights are based on correlation, for most values of $a$, the most influential nodes are change in apparel, S\&P500, and new cars. With edges directed toward the lagged variable to determine the most influenced node, PR values are shown on the right in Fig. \ref{fig5}. The most influenced nodes are change in food at home, medical, and new cars.  
The PR values are based on incoming weighted edges, and outgoing links are not considered, so any node that appears high in both lists is a node that has heavily weighted incoming links and heavily weighted outgoing links when edges are directed only toward the lead variable or only toward the lagged variable, and influence does not mean causality. Consider the variable New Cars, which is third most influential and third most influenced. Then when edges are pointed toward the lead variable, New Cars has the third highest weight of incoming edges, so it is the third most important when considering correlation with future values of other variables. New Cars also has the third highest weight of outgoing edges when edges are directed toward the lead variable, indicating that this variable is the third most important when considering correlation with past values of other variables. Correlations with past values of other variables is determined by PR when edges are directed toward the lag variable, since PR only considers incoming edges. 
Variables that switch from the top of one list to the bottom of the other, or vice versa, have a heavier weight of incoming than outgoing edges, or vice versa. For instance, Apparel has greater relations with future values of other variables but few relations with past values of other variables. Based on $a$, PR values fluctuate the most for those variables with larger standard deviation, while those with smaller standard deviation were quite insensitive to changes in $a$.

\begin{figure}
\includegraphics[width=\textwidth]{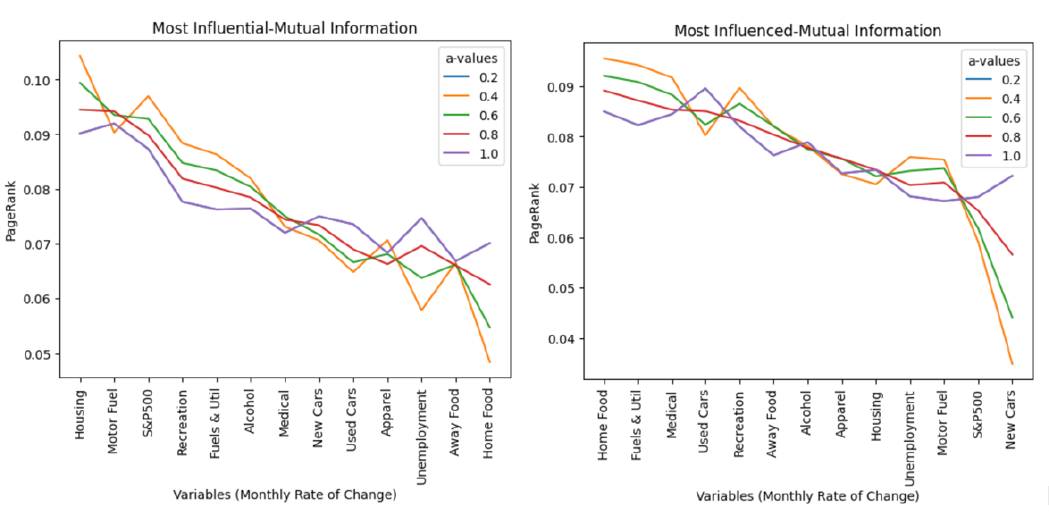} 
\caption{PR values (y-axis) per node (x-axis) for networks with edge weights based on mutual information and determined by select values of $a$ in the legend. The nodes on the x-axis are sorted by average PR. The most influential node is shown on the left and has greatest PR with edges directed toward the lead variable. The most influenced node is shown on the right and has greatest PR with edges directed toward the lagged variable.} \label{fig6}
\end{figure}

When edge weights are based on mutual information, with variables sorted by average PR, the most influential nodes are shown on the left in Fig. \ref{fig6}, and they are rate of change in housing, motor fuel, and S\&P500. These most influential nodes each have greater relations with future values of other variables. The most influenced nodes, shown on the right in Fig. \ref{fig6}, are change in food at home, fuels \& utilities, and medical, and these variables each have greater relations with past values of other variables. Food at home is the least influential and the most influenced, indicating that change in amount spent on food at home is most related to change in variables over time in previous months, though it has the lowest relational weight to other variables at a future time. Comparing the results from the mutual information network with those of the correlation network, S\&P500 is in the top three of most influential, and food at home and medical are in the top three of the most influenced. Considering the nonlinearity of the system, mutual information is likely the better metric choice.

\begin{figure}
\includegraphics[width=\textwidth]{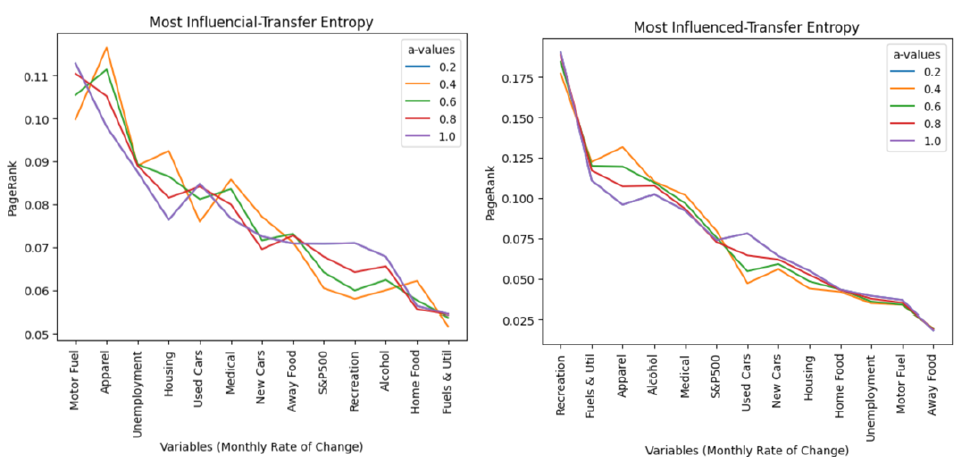} 
\caption{PR values (y-axis) per node (x-axis) for networks with edge weights based on transfer entropy and determined by select values of $a$ in the legend. The nodes on the x-axis are sorted by average PR. The most influential node is shown on the left and has greatest PR with edges directed toward the lead variable. The most influenced node is shown on the right and has greatest PR with edges directed toward the lagged variable.} \label{fig7}
\end{figure}

When edge weights are based on transfer entropy, with variables sorted by average PR, the most influential variables are shown on the left in Fig. \ref{fig7}. The most influential nodes are change in motor fuel, apparel, and unemployment, so these variables have greater relations with future values of other variables. The most influenced nodes based on transfer entropy are shown on the right in Fig. \ref{fig7}, and they are change in recreation, fuels \& utilities, and apparel, so these variables have a greater relation with past values of other variables. PR results vary less for edge weights based on transfer entropy than for edge weights based on both correlation and mutual information. There is a slight variation in PR remaining for variables with larger standard deviation, though results are most robust for the network with node relations based on transfer entropy. 

\section{Conclusion \& Limitations}
Many methods to investigate time series are concerned with prediction, and they work only with univariate data. This method is a useful way to analyze and understand multivariate time series data while removing the specific time dependence, and it can be applied to analyze other multivariate time series data in many fields. For each method used to quantify the relationships over time, the overall results are not sensitive to variations in the value of the input parameter $a$, for $0 \leq a \leq 1$, verifying the robustness of the method within each metric. However, the results are extremely sensitive to the choice of metric. Results are most insensitive when transfer entropy is used to determine the most influenced node, and transfer entropy was also the metric that had the greatest range of PR, making the ranking of variable influence easier to distinguish than those from the correlation network and the mutual information network. Transfer entropy was also the least sensitive to variables with more variation. While correlation and mutual information provided useful information, transfer entropy is definitely the best measure to quantify the relationships among economic indicators over time during the COVID-19 era. 

This method can be applied only to historical data; results are not predictive. Each relationship measure quantifies only pairwise relations with equal sample sizes. While mutual information and transfer entropy both capture nonlinear relations, correlation quantifies only linear relations.  Future work includes considering a different initial transformation to achieve better stationarity and test the difference in results, using Granger causality to determine relevant relations, and using a higher order network to extend the analysis beyond pairwise relations.

%
%
%
%

\end{document}